\documentclass[conference,a4paper]{IEEEtran}
\IEEEoverridecommandlockouts
\usepackage{cite}
\usepackage{indentfirst}
\usepackage{multirow}
\usepackage{booktabs}
\usepackage{makecell}
\usepackage{float}
\usepackage{amsmath,amssymb,amsfonts}
\usepackage{algorithmic}
\usepackage{graphicx}
\usepackage{graphicx,subfigure}
\graphicspath{{./figures/}}
\usepackage{textcomp}
\usepackage{xcolor}
\usepackage{subfigure}
\usepackage{booktabs}
\usepackage{bm}
\usepackage{amsthm}
\usepackage{epstopdf}
\def\BibTeX{{\rm B\kern-.05em{\sc i\kern-.025em b}\kern-.08em
		T\kern-.1667em\lower.7ex\hbox{E}\kern-.125emX}}

\begin{document}
	\bibstyle{IEEEtran}
	\title{Transmission With Machine Language Tokens: A Paradigm for Task-Oriented Agent Communication}

	\author{
		\IEEEauthorblockN{Zhuoran~Xiao\IEEEauthorrefmark{1},
			Chenhui~Ye\IEEEauthorrefmark{1},
			Yijia~Feng\IEEEauthorrefmark{1},
			Yunbo~Hu\IEEEauthorrefmark{1},
			Tianyu~Jiao\IEEEauthorrefmark{1},
			Liyu~Cai\IEEEauthorrefmark{1},
			and Guangyi~Liu\IEEEauthorrefmark{2}
		}

    \IEEEauthorblockA{\qquad\qquad\qquad\IEEEauthorrefmark{1}Nokia Bell Labs, Shanghai, China ~~\IEEEauthorrefmark{2}China Mobile Research Institute, Beijing, China\\
	E-mails: \{zhuoran.xiao, chenhui.a.ye, yijia.feng, yunbo.hu, liyu.cai\}@nokia-sbell.com, liuguangyi@chinamobile.com}
		}
	\maketitle

	\begin{abstract}
		The rapid advancement in large foundation models is propelling the paradigm shifts across various industries. One significant change is that agents, instead of traditional machines or humans, will be the primary participants in the future production process, which consequently requires a novel AI-native communication system tailored for agent communications. Integrating the ability of large language models (LLMs) with task-oriented semantic communication is a potential approach. However, the output of existing LLM is human language, which is highly constrained and sub-optimal for agent-type communication. In this paper, we innovatively propose a task-oriented agent communication system. Specifically, we leverage the original LLM to learn a specialized machine language represented by token embeddings. Simultaneously, a multi-modal LLM is trained to comprehend the application task and to extract essential implicit information from multi-modal inputs, subsequently expressing it using machine language tokens. This representation is significantly more efficient for transmission over the air interface. Furthermore, to reduce transmission overhead, we introduce a joint token and channel coding (JTCC) scheme that compresses the token sequence by exploiting its sparsity while enhancing robustness against channel noise. Extensive experiments demonstrate that our approach reduces transmission overhead for downstream tasks while enhancing accuracy relative to the SOTA methods.
	\end{abstract}

	\begin{IEEEkeywords}
		6G networks, large language models, agents, semantic communication, machine learning.
	\end{IEEEkeywords}

	\section{Introduction} \label{intro}
	With the rapid development of generative artificial intelligence, it is foreseeable that various vertical industries are undergoing a paradigm shift, wherein intelligent agents powered by large models are gradually replacing traditional workflows. Consequently, wireless communication systems are also expected to evolve from conventional architectures, primarily designed for human communication focusing on accurate bit-level recovery, towards novel systems that facilitate agent communication based on semantic understanding through token representations.

	The evolution of agent communication systems is expected to follow three primary directions. First, the transmission process will become highly task-coupled, meaning both the transmitter and the receiver must understand which portions of the transmitted information impact the performance of downstream tasks. Moreover, the varying significance of different components of the information must be recognized to enable efficient compression under stringent communication resource constraints. Second, future communication systems must accommodate diverse information modalities to support various application scenarios. Third, while optimized for human interpretation, existing modalities such as natural language and RGB images are sub-optimal for machine communication, both in terms of transmission efficiency and semantic precision. Hence, there is a need for a novel, universal information representation specifically tailored for efficient conveyance and comprehension by intelligent agents.

	Several studies have begun exploring more efficient communication paradigms to address these emerging needs. DeepJSCC \cite{10328187} introduced a deep learning-based joint source-channel coding framework, later adapted for specific applications such as image and video transmission \cite{9953110}. To further reduce transmission costs, task-oriented semantic communication has been proposed \cite{10333632}, enabling direct task completion without data reconstruction. With the rise of AIGC, recent works have leveraged large language models (LLMs) to extract task-relevant information, improving flexibility and generalization. For instance, \cite{10570717} used LLMs to compress textual content, while \cite{10615340} applied LLMs and diffusion models for semantic image transmission.

	Despite these promising developments, several limitations hinder applying existing methods in agent communication systems directly. Traditional JSCC methods are task-agnostic and prioritize full data reconstruction, which may not align with agent-centric objectives. While more aligned with specific downstream goals, task-oriented semantic communication approaches often involve end-to-end training tailored to a single task, making them impractical for scalable deployment in dynamic communication environments that require support for multiple tasks. Furthermore, although LLM-based semantic extraction methods show potential, their outputs remain in natural language, which may be inadequate for precisely representing critical information. Additionally, most current approaches are restricted to a single modality, limiting their applicability in multi-modal scenarios.

	This paper proposes a novel network architecture for agent semantic communication that seeks to achieve high flexibility, generalization, efficiency, and precision. This design is grounded on three key principles. First, it fully exploits the capabilities of large language models by enabling them to infer the receiver's requirements without explicitly revealing intent. Second, recognizing the inefficiency of natural language for machine communication, we propose a self-learned machine token representation by the LLM optimized for compactness and semantic expressiveness. Third, given that agents are designed for domain specific tasks rather than general-purpose understanding, the resulting machine token representations exhibit inherent sparsity, allowing further compression. Finally, the system is designed to support multi-modal inputs to ensure adaptability across diverse scenarios.

	The remainder of this paper is structured as follows. Section \ref{Problem} presents the system model and the task objective. The rationale behind the design of our network is discussed in Section \ref{network_design}. Section \ref{experiments} details the experimental setup and reports the results, providing a comparative evaluation of the proposed approach against existing state-of-the-art (SOTA) methods from multiple perspectives. Finally, Section \ref{conclusion} summarizes the main conclusions of this work.

	\begin{figure}[b!]
		\vspace{-1.5em}
		\centering
		\includegraphics[width=0.38\textwidth]{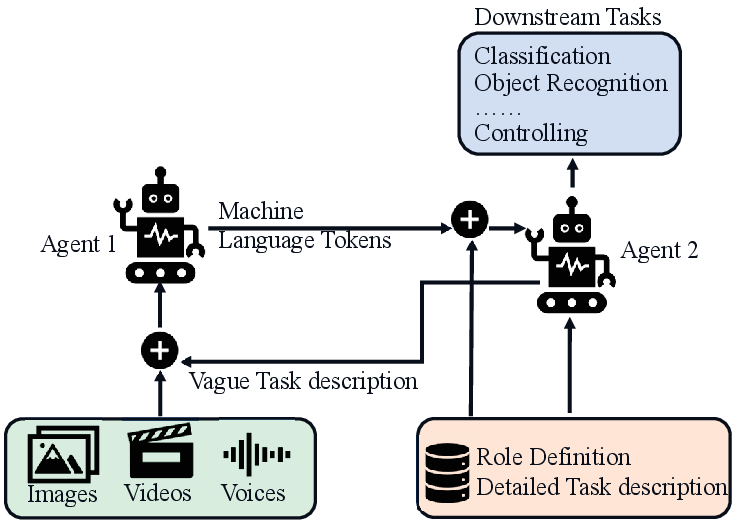}
		\caption{The system model of the proposed task-oriented agent communication.}
		\label{sys_model}
	\end{figure}

	\section{System Model And Task Description} \label{Problem}
	\subsection{System Model}
	As illustrated in Fig. \ref{sys_model}, this paper investigates a system designed to facilitate communication between two agents and the completion of specific downstream tasks. Both agents possess computational capabilities sufficient to support the deployment of at least small-scale large language models (LLMs), such as LLaMA-7B, and each agent is assigned a distinct role. Furthermore, both agents are equipped with antenna arrays and support orthogonal frequency-division multiplexing (OFDM) transmission. For clarity in the subsequent discussion, we refer to the agent that requires information to accomplish its task as the task agent and the agent that provides the necessary information as the sensor agent.

	\subsection{Task Description}
	The entire communication pipeline can be described as follows. Initially, the task agent receives a detailed description of the downstream task in natural language, denoted as $I_d$. To accomplish this task, additional information is required, which can only be obtained from the sensor agent. A straightforward approach would involve the task agent directly transmitting the detailed task description to the sensor agent, which would then return the corresponding response. However, this method is impractical in real-world systems for several reasons.

	First, the sensor agent lacks knowledge of the task agent’s role and does not possess the specialized tools available to the task agent. Second, agents are typically produced by different vendors, and due to information security and proprietary constraints considerations, they cannot expose detailed task-related data. To address these challenges, $I_d$ is first processed by a large language model (LLM), denoted by $LLM_v(\cdot)$, to perform text obfuscation and de-identification. This results in a vague or abstracted task description $I_v$, which is then sent to the sensor agent.

	The sensor agent processes $I_v$ using a language tokenizer and embedding module to obtain a language embedding $e_v$. Simultaneously, it employs a corresponding multi-modal projection module to generate a multi-modal embedding $e_m$ for its modality-specific data, such as images, videos, or audio. The embeddings $e_v$ and $e_m$ are concatenated and input into a multi-modal large language model, denoted as $LLM_t$.

	Distinct from conventional approaches, the sensor agent transmits the token embeddings from the final transformer block, referred to as machine language tokens and denoted by $T_m$, rather than natural language outputs. After transmission through a noisy wireless channel, the task agent receives a reconstructed version, $\hat{T_m}$. Subsequently, the task agent tokenizes and embeds $I_d$ to obtain the detailed task embedding $e_d$. It then concatenates $\hat{T_m}$ with $e_d$ and inputs the result into another LLM, denoted as $LLM_r$, to generate the final output $e_o$.

	Accordingly, the complete communication pipeline for the sensor agent is formulated as
	\begin{equation}
		T_m = LLM_t[cat(e_v,e_m)],
	\end{equation}
	where $cat(\cdot)$ denotes the concatenation operation. Similarly, the overall communication pipeline for the task agent is expressed as
	\begin{equation}
		e_o = LLM_r[cat(\hat{T_m},e_d)],
	\end{equation}
	where $e_d = LLM_v(I_v)$.

	\section{Learning Network Design and Motivation} \label{network_design}
	\subsection{Motivations}
	The overall architecture of the agent-based semantic communication network is depicted in Fig. \ref{nn_structure}. It is worth noting that, for clarity, we utilize images as a representation of multi-modal input. In reality, this network framework applies to any information modality. Four key motivations arise from our observations of the characteristics of large language models (LLMs) and agent tasks, which determine the proposed agent communication network structure.
	\subsubsection{Task Awareness}
	As proved in \cite{deletanglanguage}, the compression process can be interpreted as a directional collapse of information, wherein high-dimensional representations are selectively projected onto a lower-dimensional manifold aligned with task-relevant features. Nevertheless, in current research on both DeepJSCC and task-oriented semantic communication, effective methodologies remain absent to explicitly guide the direction of this information collapse. In the proposed approach, inspired by image-text alignment studies, we utilize task descriptions as directional guidance by integrating language token embeddings with multi-modal embedding modules through concatenation.
	\subsubsection{Information Representation with Machine Language Tokens}
	Given that general-purpose large language models are inherently constrained to produce outputs in natural language, existing methods for semantic communication leveraging LLMs rely on converting the original information into textual descriptions at the transmitter side. These textual representations are then transmitted and subsequently used at the receiver side to reconstruct semantically similar images or videos, which are further employed in downstream tasks.

	This paradigm is trivial and sub-optimal in terms of transmission efficiency and downstream task performance, primarily because natural language, originally designed for human communication, inherently contains a significant amount of redundancy. Moreover, natural language is fundamentally ambiguous and thus incapable of precisely representing fine-grained details in other modalities. A growing body of research on language transfer in LLMs has demonstrated that, with minimal fine-tuning, LLMs can express their internal knowledge using previously unseen languages without compromising their performance \cite{brown2020language}. Inspired by this, we propose reformulating the communication process for agent-based systems by allowing the LLM to autonomously learn a task-specific \textit{machine language} through end-to-end training with limited data.

	Unlike natural language, this machine language is represented directly through token embeddings. During the end-to-end training process, we impose constraints on the number of tokens used and simulate the noisy transmission of token embeddings over wireless channels. This enables the network to learn an information representation that is both compact and robust, tailored to the requirements of specific downstream tasks.
	\subsubsection{Prefix Information Injection}
	Previous research on prefix-tuning in LLMs has demonstrated that appending a small number of trainable tokens as a prefix to the original task instruction can significantly enhance the model's performance on downstream tasks. However, in this paradigm, the prefix remains fixed after training for a specific task. Inspired by this, we propose treating machine language tokens as trainable and non-fixed prefixes, which are concatenated before the detailed task description, serving as a novel method for knowledge injection. This approach directly tackles the downstream task without necessitating fine-tuning the task-specific model's parameters. Specifically, during training, the receiver computes the gradient of the input machine language token embeddings via backpropagation, with the gradients subsequently propagated back to the transmitter's LLM for parameter fine-tuning. Consequently, this method reduces training overhead during the training phase while ensuring the generalizability of the downstream task model during inference.

	\subsubsection{Intrinsic Sparsity Within Token Embeddings}
	Existing studies have demonstrated that, for large language models (LLMs), tokens belonging to the same domain tend to be closer in the embedding space \cite{ethayarajh2019contextual}. In order to encompass knowledge across all domains, general-purpose LLMs typically adopt high-dimensional embeddings to store and accurately distinguish representations under general tasks. However, agents are often designed to operate within more specialized vertical domains. As a result, when using general-purpose embeddings to represent such agents, the corresponding token embeddings tend to cluster within a relatively small region of the high dimensional space, reflecting a form of spatial sparsity. This observation provides theoretical support for further dimension reduction of token embeddings in general LLMs, which can help reduce transmission overhead.

	Based on the above observation, we propose further compressing the dimensions of machine language token embeddings using a neural network. Considering that the compressed token embeddings will be directly transmitted over wireless channels, we introduce simulated channel noise during the end-to-end training process to guide the network toward learning more robust compression strategies. This process is called Joint Token and Channel Coding (JTCC).

	\subsubsection{Over-The-Air Analog Transmission}
	Currently, token transmission typically relies on vector quantization to construct a vocabulary, with token indices transmitted accordingly. While effective for natural language tokens in general-purpose large language models, this approach becomes sub-optimal for the agent communication systems considered in this work, particularly under our proposed network architecture.

	To better align with the targeted scenario, we propose an analog transmission scheme that directly transmits each dimension of the compressed token embeddings as continuous signals. This design addresses several key challenges. First, index-based transmission is highly sensitive to bit errors, often causing severe semantic distortion, whereas direct transmission limits noise-induced degradation. Second, vector quantization requires large codebooks to maintain high fidelity, increasing transmission overhead, while compression and direct transmission substantially reduce the data volume. Third, by leveraging the multi-antenna and multi-carrier channel structure, the receiver can reconstruct embeddings through a single matrix operation, thus avoiding the high computational complexity and latency associated with iterative decoding. These advantages make the proposed analog strategy well-suited for efficient and robust agent communication.
		\begin{figure*}[htb!]
		\centering
		\includegraphics[width=0.85\textwidth]{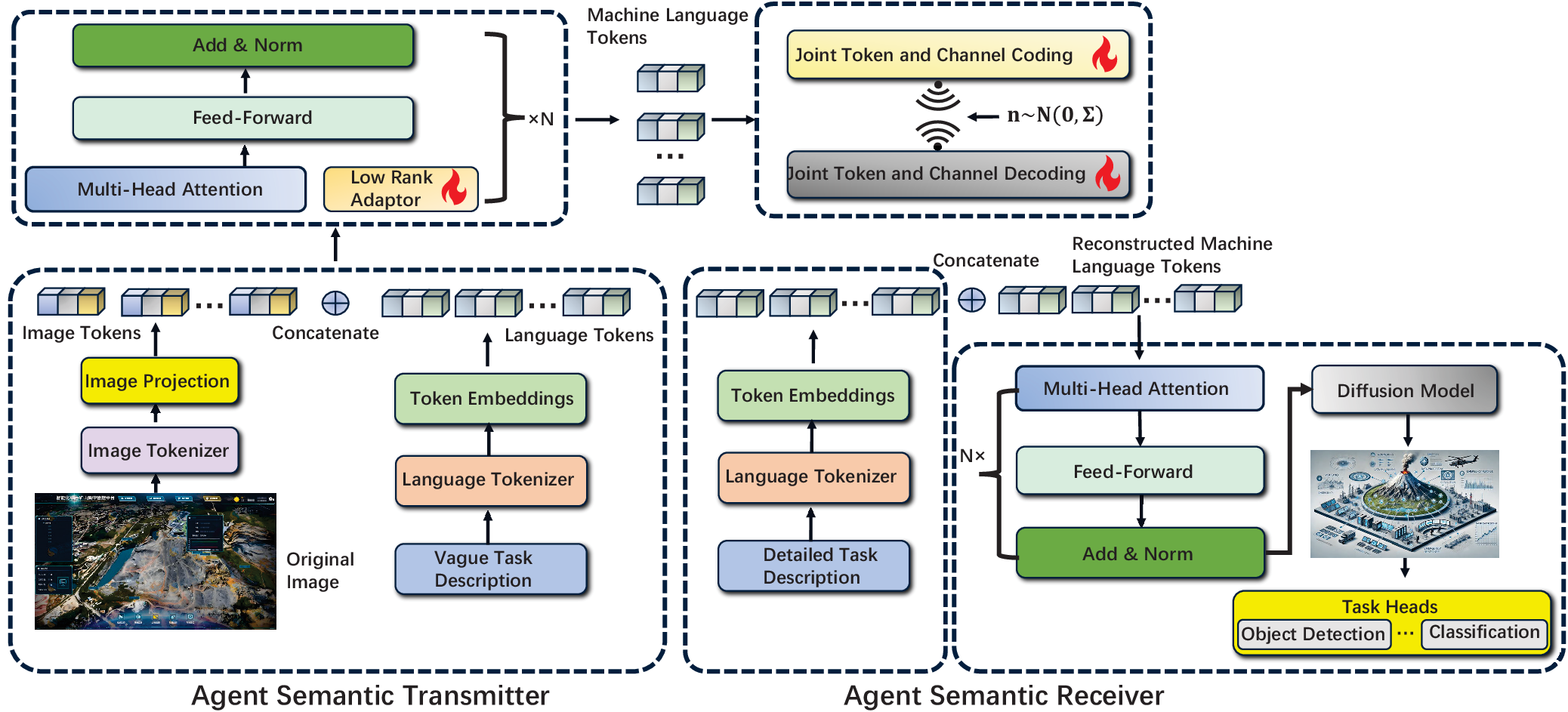}
		\vspace{-.5em}
		\caption{The network structure of the proposed task-oriented agent communication system.}
		\label{nn_structure}
		\vspace{-1.5em}
	\end{figure*}
	\subsection{Network Structure}
	In this section, we describe the network architecture shown in Fig. \ref{nn_structure} through the lens of forward propagation.

	When the semantic transmitter receives a vague description of the downstream task, it employs a pre-trained natural language tokenizer and token embedding layer to convert the original text into a sequence of token embedding vectors. We denote the embedding dimension as $L_{emb}$. Taking images as an example, the original multi-modal information is first processed by a pre-trained image tokenizer and then passed through a projection module, resulting in tokens with the same embedding dimension $L_{emb}$. Typically, an image tokenizer divides an image into patches of equal pixel count, and the image projection module, realized by vision transformers, is jointly trained with the large model during the pre-training phase. For a given modality, the number of output embedding tokens remains consistent.

	Subsequently, the multi-modal token embeddings are concatenated with the language embeddings and fed into a fine-tuned large language model composed of $N$ transformer blocks. In conventional language models, a language modeling (LM) head is typically appended after the final token of the last transformer block to predict the next token. In contrast, we propose directly using the last $K$ token embeddings from the final transformer block as machine language tokens. The complexity of the modality-specific information generally determines the number $K$. Taking images as an example, we observe that the number of tokens $K$ required to complete the downstream tasks is often less than 10\% of the number of original image tokens.

	\begin{figure}[b!]
		\centering
		\vspace{-1.5em}
		\includegraphics[width=0.42\textwidth]{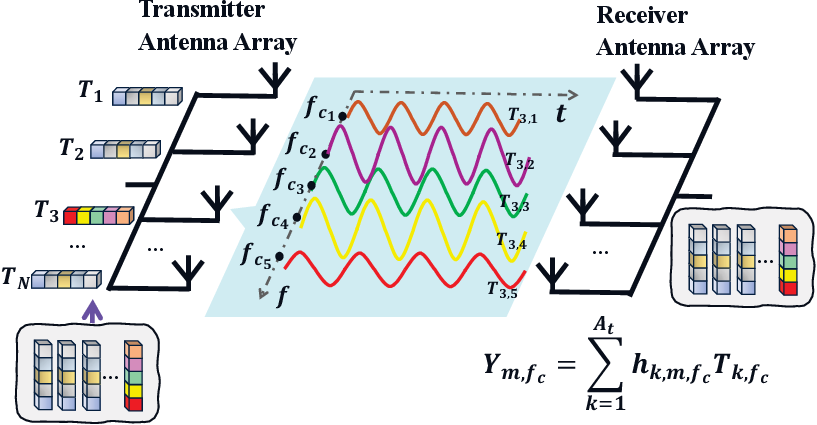}
		\caption{The propose orthogonal frequency division anolog transmission (OFAT) scheme.}
		\label{antenna}
	\end{figure}

	Afterward, the output machine language tokens are passed through a joint token and channel coding module, which reduces the dimensionality to $L_t$. Thus, the overall transmission requires sending a matrix of dimension $\mathbf{T} \in \mathbb{R}^{K \times L_t}$ over the air. We introduce an innovative MIMO transmission scheme based on Orthogonal Frequency Division Anolog Transmission (OFAT) to enhance transmission efficiency. Let $A_t$ denote the number of transmitting antennas, $A_r$ the number of receiving antennas, $K_c$ the number of system subcarriers, and $\mathbf{H} \in \mathbb{R}^{A_t \times A_r \times K_c}$ the channel matrix between the transmitter and receiver. As shown in Fig. \ref{antenna}, by modulating the values of each dimension of the tokens using amplitude modulation across different antennas and subcarriers, it is possible to transmit $K_c \times A_t$ token embedding elements within a single symbol period. Specifically, for the $f_c$th subcarrier, the transmitted vector from the transmitting array is denoted as $\mathbf{T}_{i,f_c}$. Then, the received signal of the $m$th antenna element $\mathbf{Y}_{m,f_c}$ can be written as
	\begin{equation}
		\mathbf{Y}_{m,f_c} = \sum_{k=1}^{A_t} h_{k,m,f_c}T_{k,f_c} ,
	\end{equation}
	where $h_{k,m,f_c}$ denotes the channel response between the $k$-th transmitting antenna on subcarrier $f_c$ and the $m$-th receiving antenna. If $A_r \ge A_t$ is typically the case in most practical scenarios, as receivers are often task-executing agents with superior hardware configurations, $\mathbf{T}$ can be directly reconstructed. Otherwise, compressed sensing methods can approximate $\mathbf{T}$.

	After the receiver recovers $\textbf{T}$, it passes through a joint token and channel decoding module to recover the original machine language tokens. These tokens are then concatenated with embedding the detailed task-related language description, which serves as a prefix and is fed into the receiver-side large language model (LLM). The receiver-side LLM can then either directly solve the downstream task through natural language output or generate an image with the same transmitted semantic information using a model such as a diffusion model. This image can be used to reuse the original downstream task module to address the initial task.
	\subsection{Training Process}
	The entire network is trained in an end-to-end manner. At the transmitter side, since the large model primarily performs the transformation from natural language representations to machine-interpretable forms without the need to acquire new knowledge, we propose to apply the LoRA (Low-Rank Adaptation) method by introducing a small adaptor only to the Key \newpage \noindent and Value matrices of the original transformer layers. This significantly reduces the number of trainable parameters to approximately one-thousandth of the original model size.

	Additionally, a token coding module and a token decoding module are jointly trained to achieve dimensionality compression of the token embeddings. These two modules typically adopt a symmetric autoencoder architecture.

	It is important to note that, in order to simulate the noise introduced by transmission over the air interface, the gradient cannot be directly propagated through the noisy channel. To address this challenge, we approximate the transmitter-side gradients by directly using the gradients computed from the received token embeddings at the receiver side, thereby enabling effective end-to-end optimization.

	\section{Experiments} \label{experiments}
	\subsection{Dataset generation and Parameters setting}
\begin{figure}
		\vspace{.5em}
		\centering
		\subfigure[A sample image and questions from CLEVR.]{\label{phasechange}
			\includegraphics[width=0.3\textwidth]{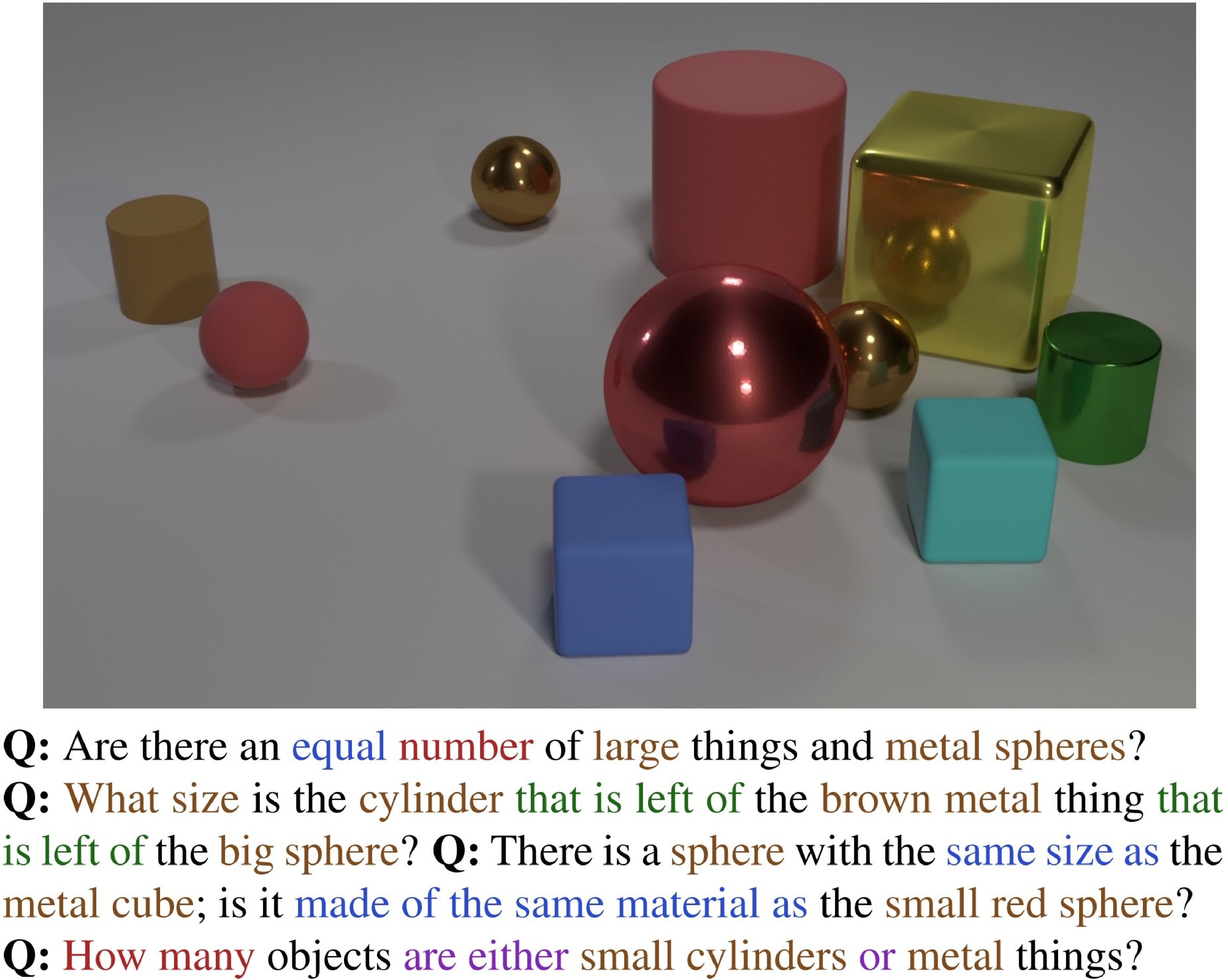}
		}
		\centering
		\subfigure[Examples from the GQA dataset for visual reasoning and compositional question answering.] {\label{phasesimulation_tanh}
			\includegraphics[width=0.3\textwidth]{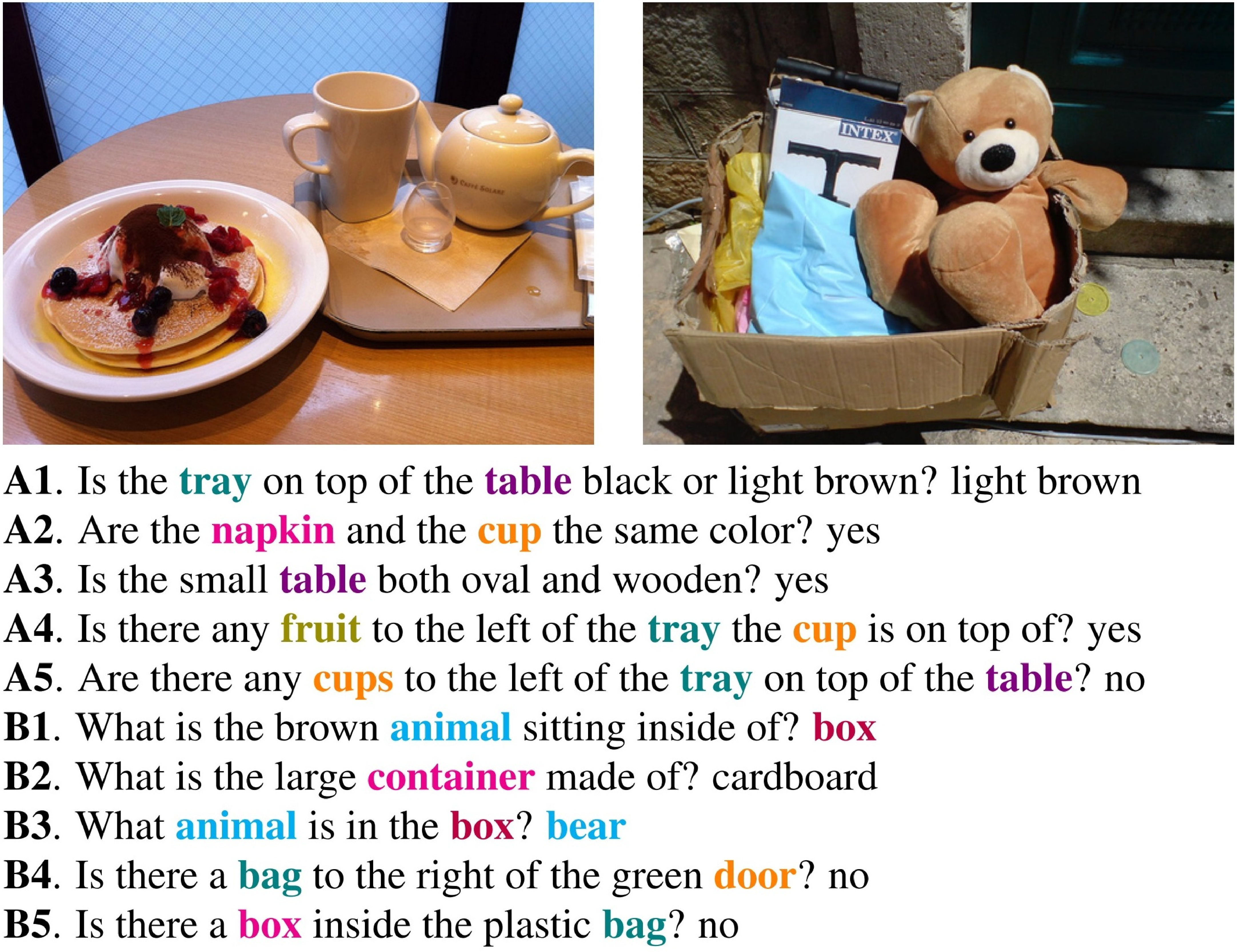}
		}
		\caption{Illustration of the datasets used in the experiment part.}
		\vspace{-.5em}
		\label{datasets}
		\vspace{-1.5em}
	\end{figure}

	We select the CLEVR \cite{8099698} and GQA \cite{hudson2018gqa} datasets, both open source benchmarks for visual question answering, to comprehensively evaluate our methods, as illustrated in Fig. \ref{datasets}. These datasets are chosen for three main reasons. First, their questions are script-generated with metadata such as color and spatial position, allowing easy construction of vague task descriptions (e.g., prompting an LLM with “This is a question related to colors in the image”). Second, answering these questions typically requires only localized visual information, aligning with real-world scenarios like intelligent manufacturing where minimal image data suffices. Third, both datasets emphasize reasoning skills, consistent with the goals of task-oriented communication, where complex tasks are implicitly solved during the communication process.

	The CLEVR training set contains 70,000 images with 10 questions each, and the test set includes 15,000 images; we adopt the same sizes for GQA to ensure fair comparison. LLaVA-V1.5-7B and LLaMA2-7B serve as the base models for the transmitter and receiver agents, respectively, selected based on accuracy, training time, and inference efficiency. LoRA is applied by inserting low-rank adapters into the Key and Value matrices of transformer blocks. The transmitter encodes information into 5 tokens, and a simple fully connected network is used in the Joint Token Compression and Communication (JTCC) module to compress token embeddings from 4096 to 256 dimensions.

	\subsection{Benchmarks}
	We select four comparison methods as benchmarks to evaluate the proposed approach regarding transmission overhead, task accuracy, and robustness.
	\begin{itemize}
		\item \textbf{Benchmark 1}: Transmitting raw images and employing the original LLaVA-V1.5-7B model to answer the questions. This benchmark reflects the baseline task accuracy without any pre-processing.
		\item \textbf{Benchmark 2}: Reconstructing the original image using DeepJSCC, followed by question answering with LLaVA-V1.5-7B. This setup evaluates the impact of DeepJSCC on transmission overhead and task accuracy.
		\item \textbf{Benchmark 3}: Utilizing LLaVA-V1.5-7B as the transmitter to summarize the transmitted information in natural language and using LLaMA2-7B to generate answers. This benchmark shows the effectiveness of using machine language tokens instead of natural languages.
		\item \textbf{Benchmark 4}: Identical to the proposed method, except without incorporating the JTCC module. This benchmark is designed to isolate and evaluate the contribution of the JTCC component.
	\end{itemize}

	\subsection{Experimental Results}
\begin{figure}
		\centering
		\includegraphics[width=0.32\textwidth]{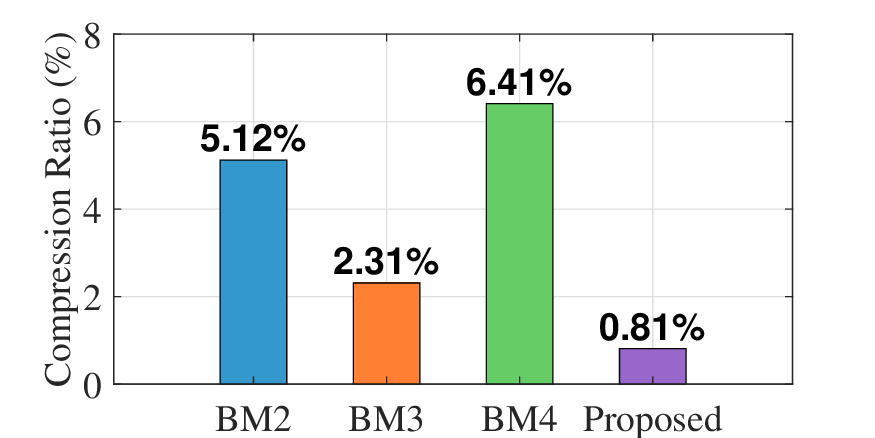}
		\vspace{-1em}
		\caption{Compression ratio relative to the original JPG images.}
		\label{compression_ratio}
		\vspace{-1.4em}
	\end{figure}

	Fig. \ref{compression_ratio} illustrates the comparison of compression rates for original images across different schemes in the experiment. Specifically, we calculated the average storage size of JPG images from the CLEVR and GQA datasets and compared it with the transmission overhead required by four methods to determine the compression ratio. Among these methods, DeepJSCC adopts the same architecture as described in [1], while benchmark 3 estimates the number of characters needed to describe image information textually, assuming UTF-8 encoding. For benchmark 4, the token embedding dimension is set to 4096, consistent with the LLaMA series of models, whereas in the proposed method, we reduce the dimensionality to 512. The experimental results demonstrate that our proposed approach significantly reduces transmission overhead under the given experimental settings.

	\begin{figure}
		\centering
		\subfigure[CLEVR dataset]{
			\begin{minipage}[c]{0.24\textwidth}
				\centering
				\includegraphics[width=1\textwidth]{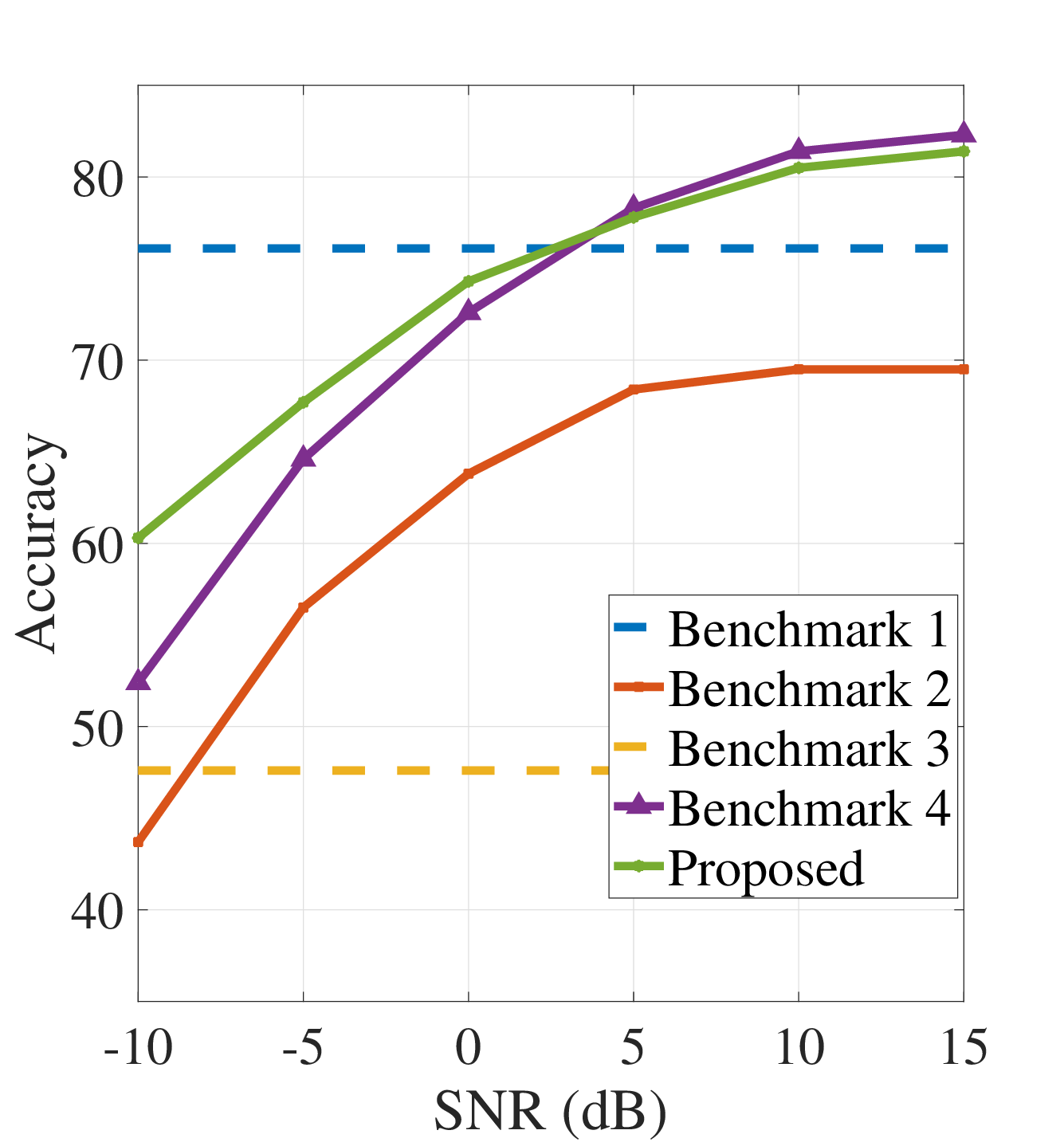}\vspace{-1em}
				\label{clevr_1}
			\end{minipage}
		}
		\hspace{-0.6cm}
		\subfigure[GQA dataset]{
			\begin{minipage}[c]{0.24\textwidth}
				\centering
				\includegraphics[width=1\textwidth]{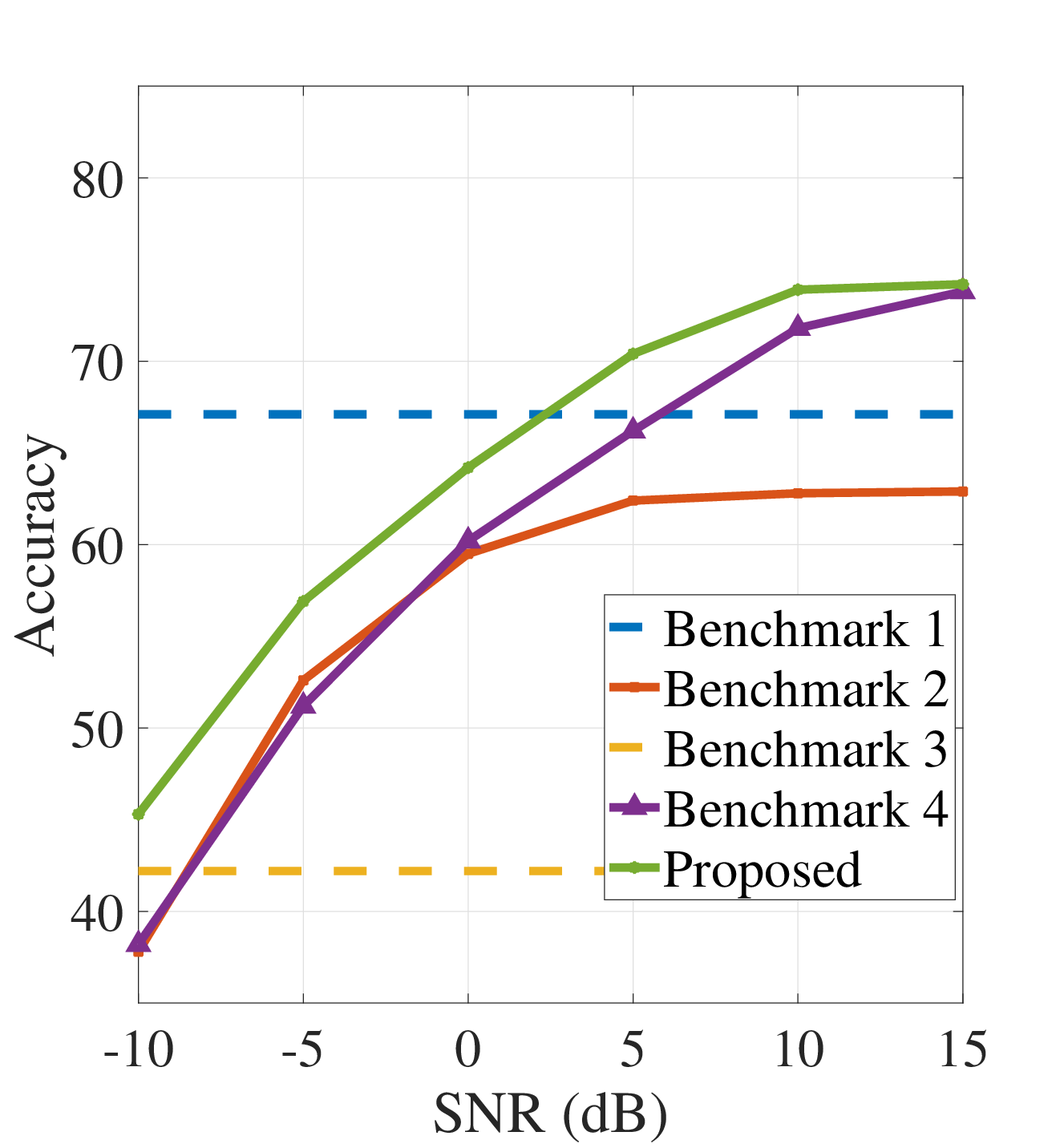}\vspace{-1em}
				\label{gqa_1}
			\end{minipage}
		}
		\vspace{-1.5em}
		\caption{The accuracy of downstream tasks across different methods with varying transmission SNR.}
		\vspace{-1.7em}
		\label{exp1}
	\end{figure}

	\begin{figure}[htb!]
		\centering
		\subfigure[CLEVR dataset]{
			\begin{minipage}[c]{0.24\textwidth}
				\centering
				\includegraphics[width=1\textwidth]{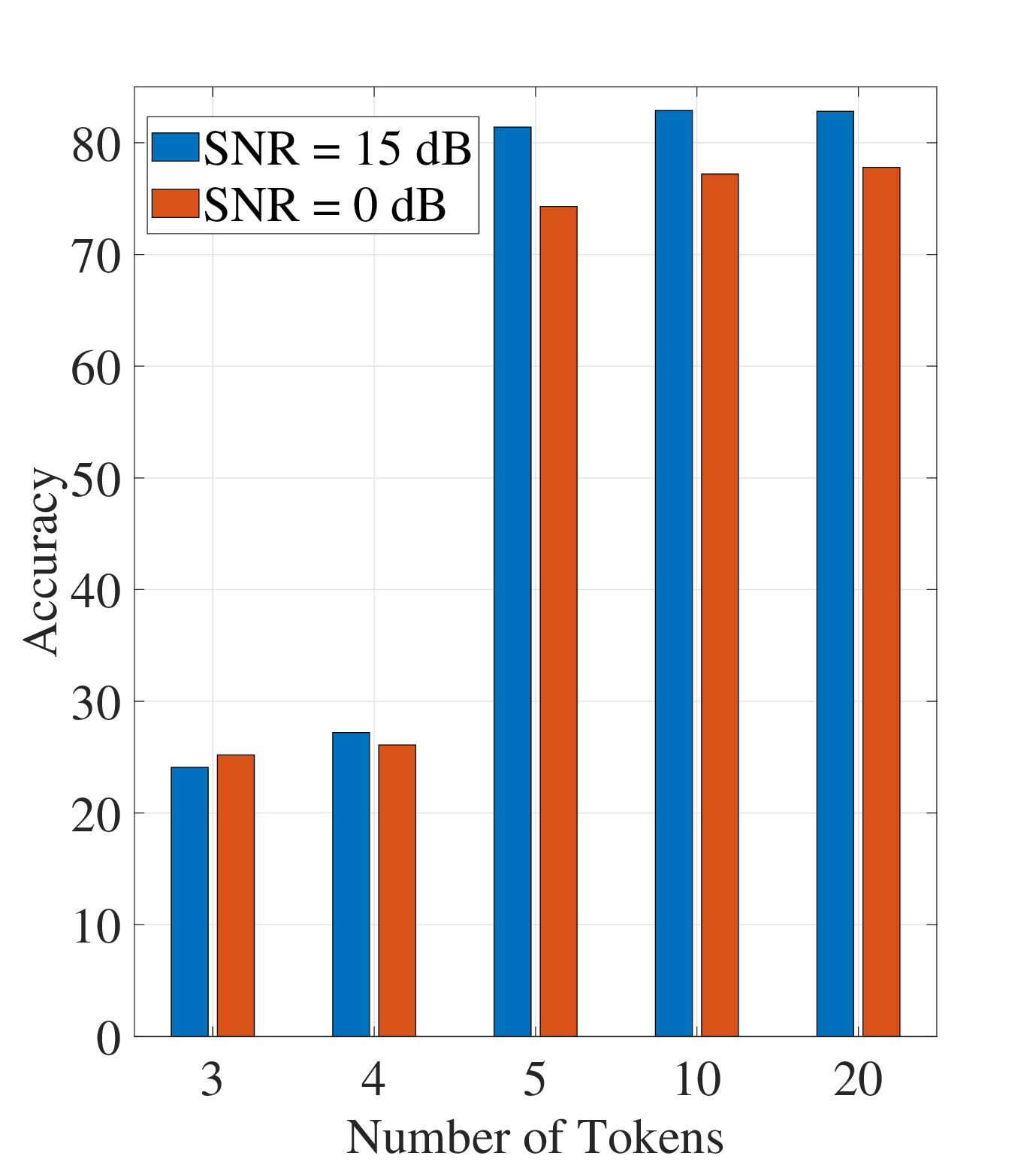}\vspace{-1em}
				\label{different_velocity_simulation}
			\end{minipage}
		}
		\hspace{-0.6cm}
		\subfigure[GQA dataset]{
			\begin{minipage}[c]{0.24\textwidth}
				\centering
				\includegraphics[width=1\textwidth]{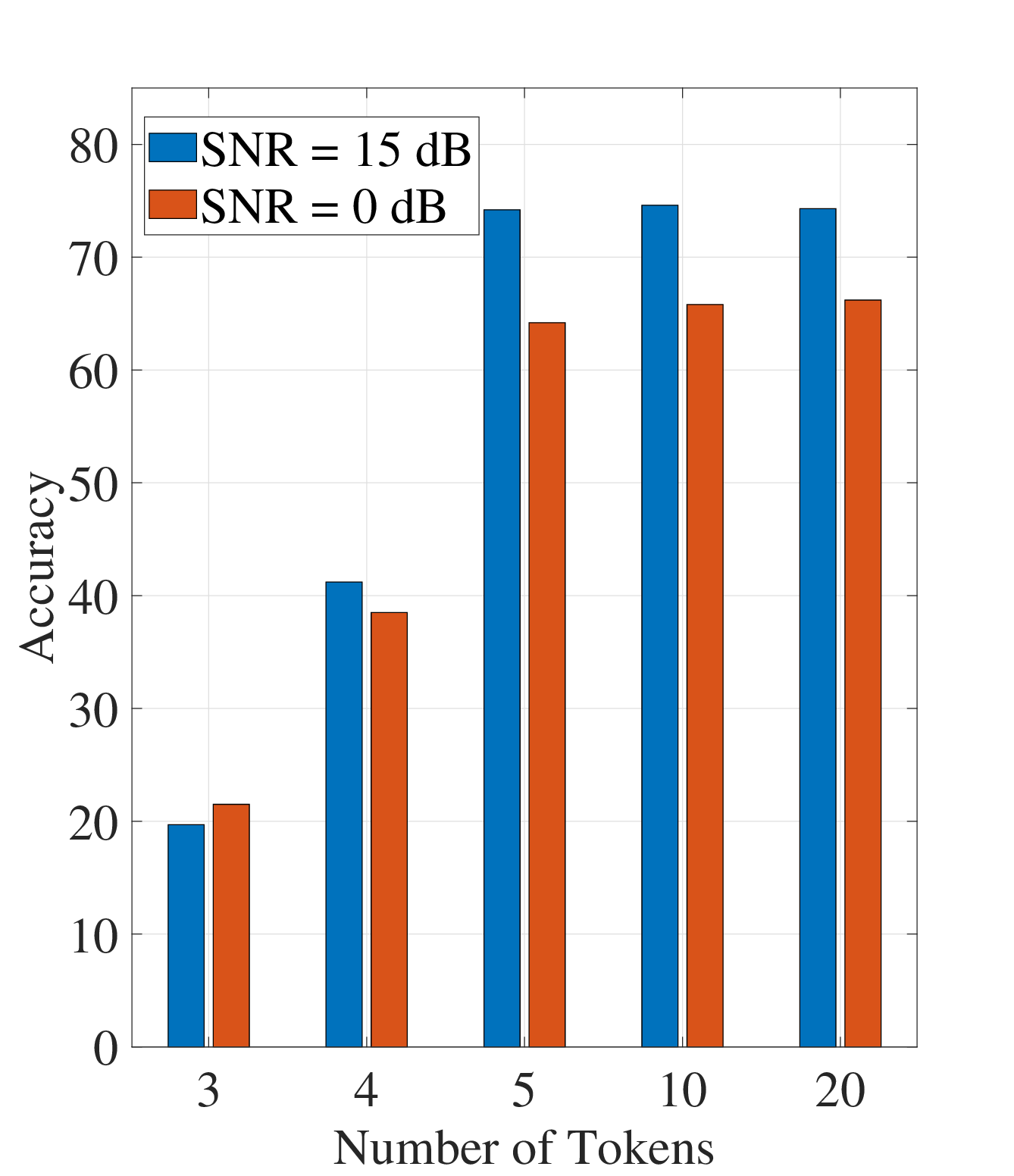}\vspace{-1em}
				\label{different_velocity_real}
			\end{minipage}
		}
		\vspace{-1.5em}
		\caption{The accuracy of downstream tasks for the proposed methods with a varying number of tokens used for transmission.}
		\vspace{-1.8em}
		\label{exp2}
	\end{figure}

	Fig. \ref{exp1} compares the downstream task accuracy of different methods across various SNRs, showing consistent trends on both datasets. Benefiting from end-to-end training on machine language tokens, the proposed method outperforms Benchmark 1 at high SNRs. In contrast, existing DeepJSCC approaches, focused solely on raw image reconstruction, fail to surpass Benchmark 1 at high SNRs and degrade significantly at low SNRs. Due to the difficulty of expressing image-related reasoning tasks clearly in natural language, Benchmark 3 performs poorly across both datasets. With the JTCC module compressing token embeddings, the proposed method achieves performance close to Benchmark 4 at high SNRs while maintaining robustness under low SNR conditions. These results demonstrate the effectiveness and robustness of the proposed method.

	Fig. \ref{exp2} investigates the influence of the number of tokens on downstream task performance under two representative SNR conditions, namely 0 dB and 15 dB. The results reveal a critical threshold phenomenon: for the two image reasoning datasets considered, five tokens are sufficient to capture all the information required for successful reasoning. Increasing the number of tokens beyond this point yields no further improvement in performance due to internal redundancy, although a marginal gain in robustness can be observed. In contrast, reducing the number of tokens below this critical threshold leads to an abrupt loss of information, resulting in substantial performance degradation in downstream tasks. These findings suggest that, for practical deployment, it is crucial to adaptively select the number of tokens based on task-specific requirements to achieve optimal transmission efficiency.

	\section{Conclusions} \label{conclusion}
	This paper introduces a pioneering paradigm for task-oriented agent communication. The uniquely architected agent semantic transmitter and receiver were designed to extract critical information and represent it using an exceptionally compact set of token embeddings through end-to-end training on domain-specific tasks. Furthermore, by incorporating a specially devised joint token and channel coding module alongside an over-the-air analog transmission mechanism, the proposed approach significantly reduces both transmission overhead and latency without compromising the accuracy of downstream tasks. The experimental outcomes affirm the efficacy and robustness of the proposed methodology, underscoring its potential to redefine future paradigms in agent communication systems.

	\section*{Acknowledgment}
	This work was supported in part by National Key Research and Development Program of China under Grant 2024YFE0200600.

	\bibliographystyle{IEEEtran}
	\bibliography{bibfile}
\end{document}